\documentclass[doublecol]{epl2}
\usepackage{graphicx,epsfig}
\usepackage{amsmath}
\date{\today}

\newcommand{\cO}{{\mathcal O}}
\newcommand{\cZ}{{\mathcal Z}}

\title{Revision of the fractional exclusion statistics}
\author{Drago\c s-Victor Anghel}
\institute{Department of Theoretical Physics, National Institute for
  Physics and Nuclear Engineering--''Horia Hulubei'', Str. Atomistilor
  no.407, P.O.BOX MG-6, Bucharest - Magurele, Romania}

\abstract{I discuss the concept of fractional exclusion statistics 
(FES) and I show that in order to preserve the thermodynamic consistency 
of the formalism, the exclusion statistics parameters should change if the 
species of particles in the system are divided into subspecies. 
Using a simple and intuitive model I deduce the general equations that 
have to be obeyed by the exlcusion statistics parameters in any FES 
system.}

\begin{document}
\maketitle

\section{Introduction}
\label{intro}

In Ref. \cite{PhysRevLett.67.937.1991.Haldane} Haldane introduced the 
fruitful concept of fractional exclusion statistics (FES). Although many 
authors analyzed the physical properties of FES systems and the microscopic 
reasons for the manifestation of this type of statistics 
(see \cite{PhysRevLett.73.922.1994.Wu,PhysRevLett.81.489.1998.Carmelo,PhysRevLett.72.600.1994.Veigy,JPhysB33.3895.2000.Bhaduri,PhysRevB.60.6517.1999.Murthy,NuclPhysB470.291.1996.Hansson,IntJModPhysA12.1895.1997.Isakov,PhysRevLett.73.3331.1994.Murthy,PhysRevLett.74.3912.1995.Sen,PhysRevLett.86.2930.2001.Hansson,JPA40.11255.2007.Comtet,PhysRevLett.80.1698.1998.Iguchi,PhysRevB.61.12757.2000.Iguchi,PhysRevLett.85.2781.2000.Iguchi} and references therein, just as examples), 
there are important properties that have been overlooked. In Ref. 
\cite{JPhysA.40.F1013.2007.Anghel} I proved that if the mutual exclusion 
statistics parameters (see below the definitions) are defined in the typical 
way (e.g. like in 
\cite{PhysRevLett.67.937.1991.Haldane,PhysRevLett.73.922.1994.Wu}), then 
the thermodynamics of the system is inconsistent. 
To restore the thermodynamics, I conjectured in the same paper that any of the 
mutual exclusion statistics parameters should be proportional to the 
dimension of the space on which it acts. 

In another paper 
\cite{PhysLettA.372.5745.2008.Anghel} I showed that FES is 
manifesting in general in systems of interacting particles and 
the calculated exclusion statistics parameters have indeed the properties 
conjectured in \cite{JPhysA.40.F1013.2007.Anghel}. 

In this letter I analyze the basic properties of the mutual exclusion 
statistics parameters based on simple, general arguments and I show that 
the conjectures introduced in \cite{JPhysA.40.F1013.2007.Anghel} are, simply, 
necessary conditions for the logical consistency of the formalism. This 
is not surprising, since the inconsistency of the thermodynamics proved
in Ref. \cite{JPhysA.40.F1013.2007.Anghel} could have been only a consequence 
of an unconsistent undelying physical model. 

\section{A simple model\label{simple_mod}}

\begin{figure}[t]
\begin{center}
\includegraphics[width=70mm]{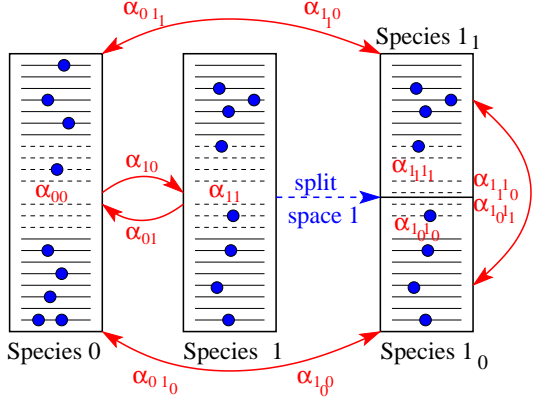}
\end{center}
\caption{(Color online) In the total system, formed of two species of 
particles, 0 and 1, the species 1 is splited into two sub-species, $1_0$ and 
$1_1$. This implies a redefinition of the exclusion statistics parameters, 
which change from the set $\tilde\alpha_{00},\tilde\alpha_{01},\tilde\alpha_{10},\tilde\alpha_{11}$, 
of the original system, into the set $\tilde\alpha_{00}',\tilde\alpha_{01_0},\tilde\alpha_{01_1},\tilde\alpha_{1_0 0},\tilde\alpha_{1_1 0},\tilde\alpha_{1_0 1_0},\tilde\alpha_{1_0 1_1},\tilde\alpha_{1_1 1_0},\tilde\alpha_{1_1 1_1}$, of the system after splitting species 1.}
\label{split}
\end{figure}

Let us assume that we have a system formed of only two species of 
particles, 0 and 1, like in Fig. \ref{split}. We 
denote the exclusion statistics parameters of this system by 
$\tilde\alpha_{00},\tilde\alpha_{01},\tilde\alpha_{10}$ and $\tilde\alpha_{11}$, and we start in 
the standard way \cite{PhysRevLett.67.937.1991.Haldane,PhysRevLett.73.922.1994.Wu} by writing the total number of configurations corresponding to $N_0$ 
particles of species 0 and $N_1$ particles of species 1 as 
\begin{equation}
W_{\{0,1\}} = \prod_i^{\{0,1\}}\frac{\left[G_i+N_i-1-\sum_j^{0,1}\tilde\alpha_{ij}
(N_j-\delta_{ij})\right]!}{N_i!\left[G_i-1-\sum_j^{0,1}\tilde\alpha_{ij}
(N_j-\delta_{ij})\right]!} , \label{W01}
\end{equation}
where $G_0$ and $G_1$ are the number of single-particle states corresponding 
to the two species of particles. We recall here that the physical 
interpretation of the exclusion statistics parameters is that at the variations 
$\delta N_0$ and $\delta N_1$ of the particle numbers $N_0$ and $N_1$, the 
number of single-particle states available for the two species changes by 
$\delta G_0=-\tilde\alpha_{00}\delta N_0-\tilde\alpha_{01}\delta N_1$ and 
$\delta G_1=-\tilde\alpha_{10}\delta N_0-\tilde\alpha_{11}\delta N_1$. 

If all the $G_0$ states have the same 
energy, say $\epsilon_0$, and all the $G_1$ states have the energy 
$\epsilon_1$, we may write the grandcanonical partition function of the 
system as 
\begin{equation}
\cZ_{\{0,1\}}=W_{\{0,1\}} \prod_i^{\{0,1\}} e^{\beta N_i(\mu_i-\epsilon_i)} , \label{Z01}
\end{equation}
where $\beta\equiv (k_BT)^{-1}$, $\mu_0$ and $\mu_1$ are the chemical potentials 
of the two species of particles, and $T$ is the temperature, common to both 
species. 

To calculate the thermodynamics of the system, we assume that all the 
numbers involved in our problem are very big, i.e. 
$N_0$, $N_1$, $G_0-1+\tilde\alpha_{00}-\tilde\alpha_{00}N_0
-\tilde\alpha_{01}N_1$, and $G_1-1+\tilde\alpha_{11}-\tilde\alpha_{10}N_0
-\tilde\alpha_{11}N_1$ are much bigger than 1. 
Maximizing $\cZ$--by calculating its logarithm and using the Stirling 
approximation--we obtain the maximum probability populations, which 
are given by the system of equations \cite{PhysRevLett.73.922.1994.Wu}
\begin{subequations} \label{Swn1}
\begin{eqnarray}
(1+w_0)\left(\frac{w_0}{1+w_0}\right)^{\tilde\alpha_{00}}
\left(\frac{w_1}{1+w_1}\right)^{\tilde\alpha_{10}} &=& e^{\beta(\epsilon_0-\mu_0)} 
\label{Sw11} \\
(1+w_1)\left(\frac{w_0}{1+w_0}\right)^{\tilde\alpha_{01}}
\left(\frac{w_1}{1+w_1}\right)^{\tilde\alpha_{11}} &=& e^{\beta(\epsilon_1-\mu_1)}
\label{Sw12} \\
(w_0+\tilde\alpha_{00})N_0+\tilde\alpha_{01}N_1 &=& G_0 
\label{Sn11} \\
\tilde\alpha_{10}N_0+(w_1+\tilde\alpha_{11})N_1 &=& G_1 
\label{Sn12}
\end{eqnarray}
\end{subequations}
%

%

\subsection{Changing the number of species}

Nevertheless, for large systems like the ones analysed above, we can 
split any of the two species of particles into 
subspecies and obtain a thermodynamically equivalent system (I shall 
prove this below). 
So let us we split for example species 1 into the 
subspecies $1_0$ and $1_1$, of dimensions $G_{1_0}$ and $G_{1_1}$. 
In this way we describe the total system as consisting of 
the species 0, $1_0$, and $1_1$, of particle numbers $N_0$, $N_{1_0}$, and 
$N_{1_1}$, in Hilbert spaces of dimensions $G_0$, $G_{1_0}$, and 
$G_{1_1}$. Obviously, 
\begin{equation}\label{N1G1split}
N_{1_0}+N_{1_1}=N_1\qquad {\rm and}\qquad G_{1_0}+G_{1_1}=G_1.
\end{equation}
We denote the exclusion statistics parameters of the ``new'' system like 
in Fig. \ref{split}, namely $\tilde\alpha_{00}'$, $\tilde\alpha_{01_0}$, $\tilde\alpha_{01_1}$, $\tilde\alpha_{1_00}$, $\tilde\alpha_{1_10}$, $\tilde\alpha_{1_01_0}$, $\tilde\alpha_{1_01_1}$, $\tilde\alpha_{1_11_0}$, $\tilde\alpha_{1_11_1}$. 
%
To obtain the consistency relations for the new exclusion statistics 
parameters, first we use the fact that the variations $\delta N_1$ and 
$\delta G_0$ may be written as 
$\delta N_1=\delta N_{1_0}+\delta N_{1_1}$ and 
$\delta G_0=-\tilde\alpha_{00}\delta N_0-\tilde\alpha_{01}\delta N_1=-\tilde\alpha'_{00}\delta N_0-\tilde\alpha_{01_0}\delta N_{1_0}-\tilde\alpha_{01_1}\delta N_{1_1}$. From these identities we obtain 
\begin{subequations} \label{relalphas}
\begin{eqnarray}
&& \tilde\alpha'_{00}=\tilde\alpha_{00}, \label{a00p} \\
&& \tilde\alpha_{01}=\tilde\alpha_{01_0}=\tilde\alpha_{01_1} \label{a01ab} 
\end{eqnarray}
by setting $\delta N_1=\delta N_{1_0}=\delta N_{1_1}=0$, 
$\delta N_0=\delta N_{1_0}=0$ or $\delta N_0=\delta N_{1_1}=0$. 
Then, writing the variation of $G_1$ as 
$\delta G_1=\delta G_{1_0}+\delta G_{1_1}$, and using the general 
expressions 
$\delta G_1=-\tilde\alpha_{10}\delta N_0-\tilde\alpha_{11}(\delta N_{1_0}
+\delta N_{1_1})$, $\delta G_{1_0}=-\tilde\alpha_{1_00}\delta N_0
-\tilde\alpha_{1_01_0}\delta N_{1_0}-\tilde\alpha_{1_01_1}\delta N_{1_1}$, and 
$\delta G_{1_1}=-\tilde\alpha_{1_10}\delta N_0-\tilde\alpha_{1_11_0}
\delta N_{1_0}-\tilde\alpha_{1_11_1}\delta N_{1_1}$,  
we obtain 
\begin{eqnarray}
&& \tilde\alpha_{10} = \tilde\alpha_{1_00}+\tilde\alpha_{1_10} \label{101a1b} \\
&& \tilde\alpha_{11} = \tilde\alpha_{1_01_0}+\tilde\alpha_{1_11_0} 
= \tilde\alpha_{1_01_1}+\tilde\alpha_{1_11_1} 
\label{111a1b}
\end{eqnarray}
\end{subequations}
by setting the independent fluctuations $\delta N_0$, 
$\delta N_{1_0}$, and $\delta N_{1_1}$ to zero in proper order.

Now we write the total number of configurations in the system, 
considering species $1_0$ and $1_1$ as distinct, 
\begin{equation}
W_{\{0,1_0,1_1\}} = \prod_i^{\{0,1_0,1_1\}}\frac{\left[G_i+N_i-1
-\sum_j^{\{0,1_0,1_1\}}\tilde\alpha_{ij}(N_j-\delta_{ij})\right]!}
{N_i!\left[G_i-1-\sum_j^{\{0,1_0,1_1\}}\tilde\alpha_{ij}
(N_j-\delta_{ij})\right]!} , \label{W01011}
\end{equation}
and we compare $\log W_{\{0,1\}}$ and $\log W_{\{0,1_0,1_1\}}$, within the 
approximation of large numbers. 

After some obvious simplifications, we obtain 
\begin{subequations} \label{logWs}
\begin{eqnarray}
\log W_{\{0,1\}} &=& (F_0+N_0)\log(F_0+N_0)+(F_1+N_1) \nonumber \\ 
&& \times \log(F_1+N_1)-F_0\log F_0 -F_1\log F_1 \nonumber \\ 
&& -N_0\log N_0-N_1\log N_1 , \label{W01F} \\
\log W_{\{0,1_0,1_1\}} &=& (F'_0+N_0)\log(F'_0+N_0)+(F_{1_0}+N_{1_0}) \nonumber \\ 
&& \times \log(F_{1_0}+N_{1_0}) +(F_{1_1}+N_{1_1}) \nonumber \\
&& \times \log(F_{1_1}+N_{1_1}) -F'_0\log F'_0 -F_{1_0}\log F_{1_0} \nonumber \\
&& -F_{1_1}\log F_{1_1} -N_0\log N_0-N_1\log N_1 , \label{W01011F}
\end{eqnarray}
\end{subequations}
with 
\begin{subequations} \label{Fs}
\begin{eqnarray}
F_0 &=& G_0+N_0-1+\tilde\alpha_{00}-\tilde\alpha_{00}N_0-\tilde\alpha_{01}N_1
\label{F0}\\
F_1 &=& G_1+N_1-1+\tilde\alpha_{11}-\tilde\alpha_{10}N_0-\tilde\alpha_{11}N_1
\label{F1}\\
F'_0 &=& G_0+N_0-1+\tilde\alpha_{00}-\tilde\alpha_{00}N_0-\tilde\alpha_{01_0}
N_{1_0} \nonumber \\
&& -\tilde\alpha_{01_1}N_{1_1} \label{Fp0}\\
F_{1_0} &=& G_{1_0}+N_{1_0}-1+\tilde\alpha_{1_01_0}-\tilde\alpha_{1_00}N_0 
-\tilde\alpha_{1_01_0}N_{1_0} \nonumber \\
&& -\tilde\alpha_{1_01_1}N_{1_1} 
\label{F10}\\
F_{1_1} &=& G_{1_1}+N_{1_1}-1+\tilde\alpha_{1_11_1}-\tilde\alpha_{1_10}N_0 
-\tilde\alpha_{1_11_0}N_{1_0} \nonumber \\
&& -\tilde\alpha_{1_11_1}N_{1_1} 
\label{F11}
\end{eqnarray}
\end{subequations}
But using Eqs. (\ref{N1G1split}) and (\ref{relalphas}), one can easily show 
that 
\begin{equation}\label{relFs}
F_0=F'_0\ {\rm and}\ F_{1_0}+F_{1_1}=F_1-1+\tilde\alpha_{1_01_0}
+\tilde\alpha_{1_11_1}-\tilde\alpha_{11}\approx F_1.
\end{equation}
Now notice that if $M$ is a big number and $c$ is a number betweem 0 and 1, 
then 
\begin{eqnarray}
&& \frac{M\log M-cM\log cM-(1-c)M\log [(1-c)M]}
{M\log M} \nonumber \\
&& = -c\frac{\log c}{\log M} - (1-c)\frac{\log (1-c)}{\log M}  \nonumber \\
&& = \cO(\log^{-1} M)\ll 1. \label{largeM}
\end{eqnarray}

 Therefore from Eqs. (\ref{relFs}) and (\ref{largeM}) we obtain that 
\begin{eqnarray}
\frac{\log W_{\{0,1\}}-\log W_{\{0,1_0,1_1\}}}{\log W_{\{0,1\}}} = 
\cO(\log^{-1} N)\ll 1, \label{rationWs}
\end{eqnarray}
where $N$ is a number comparable to $N_0$ and $N_1$. 
So indeed, as mentioned in the beginning of this subsection, in the limit of 
large numbers the splitting of the systems species into sub-species 
does not change the thermodynamics of the system, provided that the 
consistency conditions (\ref{relalphas}) are imposed on the $\alpha$s.

Now let us compare the equilibrium distributions of particles in the two 
descriptions of the system. If we maxmize the partition function 
$\cZ_{\{0,1_0,1_1\}}=W_{\{0,1_0,1_1\}}\prod_i^{\{0,1_0,1_1\}} e^{\beta N_i(\mu_i-\epsilon_i)}$, with respect to the populations we obtain the new system of equations 
\begin{subequations} \label{Swn2}
\begin{eqnarray}
e^{\beta(\epsilon_0-\mu_0)} &=& (1+w'_0)\left(\frac{w'_0}{1+w'_0}\right)^{\tilde\alpha_{00}}\left(\frac{w_{1_0}}{1+w_{1_0}}\right)^{\tilde\alpha_{1_00}} \nonumber \\
&&\times\left(\frac{w_{1_1}}{1+w_{1_1}}\right)^{\tilde\alpha_{1_10}}
\label{Sw21}\\
e^{\beta(\epsilon_1-\mu_1)} &=& (1+w_{1_0})\left(\frac{w'_0}{1+w'_0}\right)^{\tilde\alpha_{01}}\left(\frac{w_{1_0}}{1+w_{1_0}}\right)^{\tilde\alpha_{1_01_0}}\nonumber \\
&& \times\left(\frac{w_{1_1}}{1+w_{1_1}}\right)^{\tilde\alpha_{1_11_0}} 
\label{Sw22} \\
e^{\beta(\epsilon_1-\mu_1)} &=& (1+w_{1_1})\left(\frac{w'_0}{1+w'_0}\right)^{\tilde\alpha_{01}}\left(\frac{w_{1_0}}{1+w_{1_0}}\right)^{\tilde\alpha_{1_01_1}} \nonumber \\
&&
\times\left(\frac{w_{1_1}}{1+w_{1_1}}\right)^{\tilde\alpha_{1_11_1}} \label{Sw23}
\end{eqnarray}
\begin{eqnarray}
G_0 &=& w'_0 N_0 + \tilde\alpha'_{00}N_0+\tilde\alpha_{01_0}N_{1_0}+\tilde\alpha_{01_1}N_{1_1} \label{Sn21} \\
G_{1_0} &=& w_{1_0} N_{1_0} + \tilde\alpha_{1_00}N_0+\tilde\alpha_{1_01_0}N_{1_0}+\tilde\alpha_{1_01_1}N_{1_1}\label{Sn22} \\
G_{1_1} &=& w_{1_1} N_{1_1} + \tilde\alpha_{1_10}N_0+\tilde\alpha_{1_11_0}
N_{1_0}+\tilde\alpha_{1_11_1}N_{1_1} \label{Sn23}
\end{eqnarray}
\end{subequations}
where we used the notation $w'_0$, to distinguish  the solutions 
of the system (\ref{Swn2}) from the solutions of the system (\ref{Swn1}). 
Using the conditions (\ref{relalphas}), we can compare the 
two sets of solutions. 

We start with Eq. (\ref{Sn21}) in which we plug Eqs. (\ref{a00p}),
(\ref{a01ab}), and (\ref{N1G1split}); we 
obtain $(w'_0+\tilde\alpha_{00})N_0+\tilde\alpha_{01}N_1=G_0$, 
so we can conclude, after inspecting Eq. (\ref{Sn11}), that 
\begin{subequations}\label{relws}
\begin{equation}
w'_0 = w_0 . \label{w0pw0}
\end{equation}
To obtain a relation between $w_1$, $w_{1_0}$, and $w_{1_1}$, we add 
Eqs. (\ref{Sn22}) and (\ref{Sn23}). After some simple algebraic manipulations, 
using Eqs. (\ref{relalphas}) and (\ref{N1G1split}), we obtain 
$G_1=\tilde\alpha_{10}N_0+(w_{1_1}+\tilde\alpha_{11})N_1+(w_{1_0}-w_{1_1})N_{1_0}$, 
which should hold for arbitrary $N_{1_0}<N_1$. Comparing this result with 
Eq. (\ref{Sn12}) we conclude that 
\begin{equation}
w_1 = w_{1_0}=w_{1_1} \label{w1w1a1b}
\end{equation}
\end{subequations}

Using now Eqs. (\ref{relalphas}) and (\ref{relws}) we observe that 
Eqs. (\ref{Sw21}), (\ref{Sw22}), and (\ref{Sw23}) are reduced to the 
Eqs. (\ref{Sw11}) and (\ref{Sw11}), so the systems of equations 
(\ref{Swn1}) and (\ref{Swn2}) are indeed equivalent. 

Therefore if FES is manifesting into a system, the only physically consistent 
way of defining it is to impose on its exclusion statistics parameters 
the conditions (\ref{relalphas}). 

\section{The generalization of the simple model\label{gen_mod}}

We can extend the model of the previous section to a system of 
arbitrary number of particle species. We denote now by $N_i$ and 
$G_i$ the particle number and the dimension of the single-particle 
space that contain the species $i$, with $i=0,1,\ldots$. If we split 
any of the species, say species $j$, into a number of sub-species, 
$j_0,j_1,\ldots$, then all the parameters $\tilde\alpha_{kl}$, with 
both, $k$ and $l$ different from $j$, remain unchanged, whereas the 
rest of the parameters must satisfy the relations
\begin{subequations}\label{relalphasgen}
\begin{eqnarray}
\tilde\alpha_{ij} &=& \tilde\alpha_{ij_0}=\tilde\alpha_{ij_1}=\ldots,\ 
{\rm for\ any}\ i,\ i\ne j \label{alphaij} \\
\tilde\alpha_{ji} &=& \tilde\alpha_{j_0i}+\tilde\alpha_{j_1i}+\ldots,\ 
{\rm for\ any}\ i,\ i\ne j \label{alphaji} \\
\tilde\alpha_{jj} &=& \tilde\alpha_{j_0j_0}+\tilde\alpha_{j_1j_0}+\ldots 
\nonumber \\
&=& \tilde\alpha_{j_0j_1}+\tilde\alpha_{j_1j_1}+\ldots = \ldots
\label{alphajj}
\end{eqnarray}
\end{subequations}

\subsection{The ``extensivity'' of the mutual exclusion statistics parameters}

Notice that the property (\ref{alphaji}) of the mutual exclusion 
statistics parameters is satisfied for a given pair of species, $i$ and $j$, 
$i\ne j$, if 
$\tilde\alpha_{ji}$ satisfy the relation 
\begin{equation}
\tilde\alpha_{ji}/G_{j}=\tilde\alpha_{j_0i}/G_{j_0}=\tilde\alpha_{j_1i}/G_{j_1}
=\ldots=\alpha_{ji}, \label{alphaext}
\end{equation}
for any division of the space $G_j$, where $\alpha_{ij}$ is a constant for 
the pair $(i,j)$. In such a situation $\tilde\alpha_{ji}$ 
\textit{is proportional to the dimension of the space on which it acts}--$G_j$ 
and $G_{j_i}$ in Eq. (\ref{alphaext}); we say $\tilde\alpha_{ji}$ ``extensive'' 
\cite{JPhysA.40.F1013.2007.Anghel}. 

Let us assume that for a given system, we can find a fine enough division 
into species, such that the extensivity condition (\ref{alphaext}) is 
satisfied. Therefore we can write 
\begin{equation}
\tilde\alpha_{ij} = G_i\alpha_{ij}, \label{talphaext}
\end{equation}
and we apply the general formalism introduced in Ref. 
\cite{JPhysA.40.F1013.2007.Anghel}. 
The populations of the single-particle levels are given by the 
set of equations 
\begin{equation}
\beta(\mu_i-\epsilon_i)+\ln\frac{[1+n_i]^{1-\alpha_{ii}}}{n_i} = 
\sum_{j(\ne i)} G_{V_{j}}\ln[1+n_{j}] \alpha_{ji}
\label{inteq_for_n1}
\end{equation}
where $\mu_i$ and $\epsilon_i$, are the chemical potential 
and the energy level of species $i$ ($i=0,1,\ldots$). 

Some care should be taken with Eq. (\ref{inteq_for_n1}), since species $i$ 
of the l.h.s may be divided into sub-species and this would modify 
both sides of the equation. Therefore Eq. (\ref{inteq_for_n1}) is applicable 
without any ambiguities in the limit in which the subspecies $i$ is 
sufficiently small, so that further division would not modify the 
equation significantly. Nevertheless, in the 
thermodynamic (quasi-continuous) limit the summations are
transformed into integrals and we obtain the integral equation
\begin{equation}
\beta(\mu_i-\epsilon_i)+\ln\frac{[1+n_i]^{1-\alpha_{ii}}}{n_i}=
\int \sigma_{j}\ln[1+n_{j}]\alpha_{jj}\,dj .
\label{inteq_for_n2}
\end{equation}
where all ambiguities are removed. 

\section{Conclusions}

In this letter I deduced the general conditions necessary for the 
consistency of the fractional exclusion statistics (FES) formalism. 
In accordance with Refs. \cite{JPhysA.40.F1013.2007.Anghel,PhysLettA.372.5745.2008.Anghel,submitted.FES.RJP.Anghel}, I showed that the 
exclusion statistics parameters, $\alpha_{ij}$, are not constants, but 
they change with the species of particles in the system. The consistency 
conditions on $\alpha$s are given as Eqs. (\ref{relalphasgen}). 

A particular case for which Eqs. (\ref{relalphasgen}) are satisfied is when 
the mutual exclusion statistics parameters are proportional to the dimension 
of the space on which they act (see Eq. \ref{alphaext}), as conjectured 
in Ref. \cite{JPhysA.40.F1013.2007.Anghel}. One can eventually find in a 
physical system a fine enough coarse-graining for which Eq. (\ref{alphaext}) 
is satisfied; in such a case the most probable particle ocupation numbers 
are given by Eqs. (\ref{inteq_for_n1}) or (\ref{inteq_for_n2}). 

In Ref. \cite{PhysLettA.372.5745.2008.Anghel} I showed that general 
systems of interacting particles may be described as ideal systems with 
FES. The exclusion statistics parameters were 
calculated and it was proven that the mutual parameters obey 
Eq. (\ref{alphaext}) mentioned above. 


\end{document}